\documentclass[a4paper, amsfonts, amssymb, amsmath, reprint, showkeys, nofootinbib, twoside,superscriptaddress, floatfix]{revtex4-1}
\usepackage[english]{babel}
\usepackage[utf8]{inputenc}
\usepackage[colorinlistoftodos, color=green!40, prependcaption]{todonotes}

\usepackage{amsthm}
\usepackage{mathtools}
\usepackage{xcolor}
\usepackage{graphicx}
\usepackage[left=23mm,right=13mm,top=35mm,columnsep=15pt]{geometry} 
\usepackage{adjustbox}
\usepackage{placeins}
\usepackage[T1]{fontenc}
\usepackage{lipsum}
\usepackage{csquotes}
\usepackage[version=4]{mhchem}
\usepackage{siunitx}  
\usepackage{bm}

\usepackage[pdftex, pdftitle={Article}, pdfauthor={Author}]{hyperref} % For hyperlinks in the PDF
\DeclareSIUnit\formulaunit{f.u.}
\DeclareSIUnit\atom{atom}
\DeclareSIUnit\pbar{bar}
\DeclareSIUnit{\angstrom}{\textup{\AA}}
\newcommand{\comment}[1]{}
\usepackage{xcolor}
\usepackage{booktabs}
\usepackage{array}
\usepackage{hyperref}
\usepackage[capitalize]{cleveref}
\usepackage{svg}
\crefname{subsection}{subsection}{subsections}
\begin{document}
\title{Neural-network-enabled molecular dynamics study of \ce{HfO2} phase transitions}

\author{Sebastian Bichelmaier}
\affiliation{Institute of Materials Chemistry, TU Wien, A-1060 Vienna, Austria}
\affiliation{KAI GmbH, Europastrasse 8, A-9524 Villach, Austria}
\author{Jesús Carrete}
\affiliation{Instituto de Nanociencia y Materiales de Aragón (INMA), CSIC-Universidad de Zaragoza, E-50009 Zaragoza, Spain}
\affiliation{Institute of Materials Chemistry, TU Wien, A-1060 Vienna, Austria}
\author{Georg K. H. Madsen}
\email[Correspondence email address: ]{georg.madsen@tuwien.ac.at}%
\affiliation{Institute of Materials Chemistry, TU Wien, A-1060 Vienna, Austria}

\date{\today} 

\begin{abstract}
The advances of machine-learned force fields have opened up molecular dynamics (MD) simulations for compounds for which ab-initio MD is too resource-intensive and phenomena for which classical force fields are insufficient. Here we describe a neural-network force field parametrized to reproduce the r$^{\mathrm{2}}$SCAN potential energy landscape of \ce{HfO2}. Based on an automatic differentiable implementation of the isothermal-isobaric $(NPT)$ ensemble with flexible cell fluctuations, we study the phase space of \ce{HfO2}. We find excellent predictive capabilities regarding the lattice constants and experimental X-ray diffraction data. The phase transition away from monoclinic is clearly visible at a temperature around \SI{2000}{\kelvin}, in agreement with available experimental data and previous calculations.  Another abrupt change in lattice constants occurs around \SI{3000}{\kelvin}. While the resulting lattice constants are closer to cubic, they exhibit a small tetragonal distortion, and there is no associated change in volume. We show that this high-temperature structure is in agreement with the available high-temperature diffraction data.
\end{abstract}

%\keywords{}
\maketitle
\section{Introduction}
The atomistic modelling of structural phase transitions and the resulting changes in macroscopic material properties can have a profound impact on the development of new technologies. However, ab-initio methods have been hamstrung by the severe computational cost they incur and thus their predictive reach has been limited in practice. On the other end of the spectrum lie classical interatomic potentials. These are typically tailored to a few specific use cases for which they allow a descriptive simulation to be run in reasonable time, but prove challenging when trying to extend their applicability.

Machine-learned force fields (MLFF) are emerging as a third possibility, offering not only predictive and transferable ab-initio-like accuracy but also a reasonable computational cost. MLFFs have powered a broad variety of exploratory structural phase transition studies which were previously handicapped by resource limitations. This includes molecular dynamics (MD) simulations \cite{sivaraman_prl_21,Verdi2021,Chen_PRM23,Chtchelkatchev_JCP23} but also alternative strategies such as nested sampling \cite{Marchant_NPJCM23,Unglert_PRM23} and effective harmonic potentials (EHPs).\cite{bichelmaier_23b}

The present study is concerned with hafnia, \ce{HfO2}, which is generally thought to undergo temperature-induced phase transitions from monoclinic ($P2_1/c$) to tetragonal ($P4_2/nmc$) to cubic ($Fm\bar{3}m$) structures.\cite{wang_ACS_2006} Although the last of these phases has garnered significant attention in connection to its stability at high temperatures, the experimental and theoretical evidence regarding its cubic nature remains inconclusive. While earlier high-temperature X-ray diffraction studies report it\cite{tobase_pss_18,hong_nature_2018}, a more recent study found it to be virtually indistinguishable from a tetragonal phase.\cite{sivaraman_prl_21} 
We recently used EHPs in combination with a MLFF to study the thermal expansion and phase-transitions in \ce{HfO2}.\cite{bichelmaier_23b}  We found a $P2_1/c$-to-$P4_2/nmc$ transition at temperatures consistent with experimental results. However, while the cubic phase was found to be mechanically stable at high temperatures, it was not found to be thermodynamically stable under any applied temperature and pressure condition.\cite{bichelmaier_23b}
 The absence of cubic phases in the EHP study\cite{bichelmaier_23b} contradicts ab-initio MD (AIMD) simulations of \ce{HfO2} which indicated the appearance of the cubic phase at high temperatures \cite{fan_jpc_19}. While the disagreement between the different experimental findings and the computational results could be due to defect stabilization of the high-symmetry cubic phase \cite{kaiser_22}, the disagreement between the two purely computational approaches (EHP  and AIMD) \cite{bichelmaier_23b,fan_jpc_19} is somewhat puzzling. Both computational approaches (EHP  and AIMD) were backed by similar ab-initio methods and performed on non-defect-laden stoichiometric phases. The situations somewhat resembles that of hafnia's ``sister'' compound \ce{ZrO2}, where a study based on MLFF-backed MD simulation \cite{Verdi2021} reported finding the cubic phase at high temperatures, whereas an EHP study found it be thermodynamically unstable relative to the tetragonal phase at any realistic temperature \cite{Tolborg_CGD23}.

One possible explanation for the discrepancies in earlier studies could be differences in the interpretation of results rather than the methodologies themselves. In X-ray diffraction, distinguishing between cubic and tetragonal phases relies on the absence of peaks specific to the tetragonal phase. However, at the relevant temperatures around 3000~K, only a limited number of low-angle peaks are typically observed making it difficult to unambiguously determine the underlying symmetry.
Similarly, identifying of the underlying equilibrium structure when performing MD simulations can be challenging. It is worth pointing out that the MD studies indicating the presence of the high-temperature cubic phases in \ce{HfO2} \cite{fan_jpc_19} and \ce{ZrO2} \cite{Verdi2021} were both based on the observation of averaged lattice constants from $NPT$ simulations. Rather large fluctuations can be expected in the underlying trajectories and the conclusions could be affected by how these trajectories are interpreted. 

To elucidate why the earlier studies reach different conclusions, we perform a neural-network force field (NNFF)-backed MD study of \ce{HfO2}. We take advantage of the possibilities offered by automatic differentiation and extend the $NPT$ formalism implemented in \textsc{jax-md} \cite{jaxmd2020}, to include flexible cell shape changes, which allows for a direct study of phase transitions. We combine those developments in an extensive MD investigation of the temperature-dependent behavior of stoichiometric hafnia. We discuss the interpretation of the obtained trajectory files and find two phase transitions: one from monoclinic to tetragonal at a temperature around \SI{2000}{\kelvin} and another to a cell with a small tetragonal distortion at around \SI{3000}{\kelvin}. The findings are shown to be in agreement with the available diffraction data.

\section{Methods}

\subsection{Density Functional Theory}
The DFT data used to train the NNFF were based on the structures used for the previous EHP study of \ce{HfO2} \cite{bichelmaier_23b}. These were recalculated using the r$^2$SCAN meta-generalized gradient functional \cite{r2SCAN} within the projector-augmented-wave formalism \cite{bloechl_PRB_1994} as implemented in VASP \cite{kresse_PRB_1996,kresse_PRB_1999}. The original dataset was extended by an additional \num{200} ``flipped'' structures. Each of those is generated by extracting a sample from the existing database and randomly changing the chemical identity of some of the constituent atoms (from \ce{Hf} to \ce{O} and vice-versa), allowing for a richer sampling of e.g. \ce{Hf}-\ce{Hf} bonds.
An energy cutoff of \SI{600}{\electronvolt} and a $\Gamma$-only $k$-point mesh was used for all calculations.

\subsection{Neural-Network Force Field}
We use the \textsc{NeuralIL} architecture described in Refs.~\citenum{montes_22,carrete23}. Specifically, we choose a committee of NNFFs with five ResNet-style layers consisting of 128 : 64 : 32 : 16 : 16 neurons each and augmented with a Morse potential. The network is constructed with a cutoff radius of $r_c=\SI{5}{\angstrom}$ for the local element-specific spherical Bessel functions, an embedding dimension of $4$ and a total of $128$ basis functions. The loss function is defined as
\begin{equation}
\begin{split}
    \mathcal{L}  =&  \frac{\SI{0.1}{\electronvolt.\angstrom^{-1}}}{n_\text{at}}\sum_i^{n_\text{at}}\! \log\left[\cosh\left(\frac{\sqrt{\frac{1}{3}\sum_\alpha \Delta f_{i\alpha}^2}}{\SI{0.1}{\electronvolt.\angstrom^{-1}}}\right)\right] \\
    &+ \frac{\SI{0.1}{\electronvolt}}{{n_{\text{at}}}}\log\left[\cosh\left(\frac{\Delta E}{\SI{0.1}{\electronvolt}}\right)\right],
\end{split}
\label{eq:loss}
\end{equation}
with $\Delta E = E_{\mathrm{DFT}} - E_{\mathrm{NN}}$ and $\Delta f_{i \alpha} = f_{i \alpha; \mathrm{DFT}} - f_{i \alpha;\mathrm{NN}}$ where the indices $i$ run over the atoms in the unit cell and the $\alpha$ run over Cartesian axes. The optimization of the loss landscape described by \cref{eq:loss} was performed using the Versatile Learned Optimizer (VeLO) \cite{metz2022velo} for $21$ epochs. Compared to our previous work \cite{bichelmaier_23b} this constitutes a reduction in training time by two orders of magnitude, similar to our earlier findings for VeLO \cite{carrete23}. With this architecture we find the NNFF closely reproduces the training data, showing mean absolute errors over the validation data of \SI{5}{\milli\electronvolt\per\atom} and \SI{155}{\milli\electronvolt\per\angstrom} for energies and forces, respectively. 

\subsection{Molecular dynamics}
Flexible-cell MD calculations were performed using a modified version of \textsc{jax-md} described in detail in the next section \cite{jaxmd2020}. A $768$-atom \ce{HfO2} supercell and temperatures ranging from \SI{500}{\kelvin} to \SI{3250}{\kelvin} at ambient pressure, $p=\SI{1}{\pbar}$, were used.
A timestep of $\Delta t = \SI{.5}{\femto\second}$ was chosen after careful evaluation of convergence and the simulations covered a total of \SI{120}{\pico\second}. The coupling constants used to construct the mass-like quantities, $Q_k$, $Q_k'$ and $W_g$ were set to $\tau_T = \SI{50}{\femto\second}$ and $\tau_p=\SI{500}{\femto\second}$, respectively, ensuring efficient equilibration while preventing excessive oscillations. 

\subsection{X-ray diffraction patterns}
To obtain X-ray diffraction (XRD) patterns from the molecular dynamics trajectories we use the formalism described in the work by Zhang et al. \cite{zhang2016concept}, where the static structure factor is calculated as
\begin{equation}
    S(\mathbf{k}) = \frac{1}{n_{\mathrm{at}}}\left\langle\left|\sum_{i=1}^{n_{\mathrm{at}}}\cos\left(\mathbf{k}\cdot\mathbf{r}_i\right)\right|^2 + \left|\sum_{i=1}^{n_\mathrm{at}}\sin\left(\mathbf{k}\cdot\mathbf{r}_i\right)\right|^2\right\rangle.
    \label{eq:xrd}
\end{equation}
The data is then smoothed and consolidated using the \textsc{scikit-learn} \cite{scikit-learn} implementation of a RadiusNeighborRegressor with a radius $r_{\mathrm{XRD}}=0.5$ and a local Gaussian weighting with $\sigma=0.02$. 

\section{Flexible cell MD via automatic differentiation}
As \textsc{NeuralIL} is built on top of the \textsc{jax} framework \cite{deepmind2020jax}, it integrates well with \textsc{jax-md} \cite{jaxmd2020}, an MD library also based on \textsc{jax}. An implementation of a Nosé-Hoover chain barostat \cite{martyna_92} taking only isotropic cell changes into account was already available in \textsc{jax-md} \cite{jaxmd2020}. The barostat is based on the theory developed by Martyna, Tuckerman, Tobias and Klein (MTTK), which provides a particularly elegant formulation of the $NPT$ ensemble. In contrast to e.g. the Berendsen prescription \cite{berendsen1984}, the MTTK barostat allows sampling of the correct ensemble \cite{tuckerman2001} and thus provides physically meaningful results. Furthermore, it is less susceptible to the large-scale oscillations \cite{martyna1994} and hysteresis phenomena sometimes exhibited by systems driven by the basic Hoover barostat \cite{hoover1985,hoover1986} or even the Parrinello-Rahman barostat \cite{parrinello1981}. 

While the barostat accounting only for isotropic cell changes is sufficient for gaseous or liquid systems, flexible cell changes are necessary for the study of phase transitions in solids. Thus, the \textsc{jax-md} code was extended to incorporate the additional functionality of anisotropic or flexible cell changes \cite{YU2010294}.
We do not detail the full theory here as it is excellently described elsewhere (e.g. in  Ref.~\citenum{tuckerman2010statistical}), but limit the discussion to the issues related to implementing it in an automatic differentiation framework. While the present methodology was applied to a descriptor-based NNFF, it can be be applied equally well to one based on a message-passing NN \cite{Langer_JCP23}. The pressure tensor used to effect the time propagation of the generalized ``box momentum'' is given by
\begin{equation}
  P_{\alpha\beta}^{\mathrm{(int)}}(\mathbf{p}, \mathbf{r}) = \frac{1}{\det(\mathbf{h})}\sum_{i=1}^{n_\mathrm{at}}\bigg[\frac{p_{i\alpha}p_{i\beta}}{m_i} 
  + f_{i\alpha}r_{i\beta}\bigg] - \sigma_{\alpha\beta},
  \label{eq:pint}
\end{equation}
with $\mathbf{h}$ representing the matrix of cell vectors. $\mathbf{p}$ and $\mathbf{r}$ are $3n_\textrm{at}$-vectors containing the particle momenta and position vectors respectively. $m_i$ denote the nuclear masses and the $i$ and Greek letter indices are used like in \cref{eq:loss}. $\sigma_{\alpha\beta}$ represents a component of the stress tensor, $\bm{\sigma}$, which can be obtained through automatic differentiation of the position- and cell-dependent energy function $E\left(\mathbf{r}, \mathbf{h}\right)$:
\begin{equation}
    \bm{\sigma} = \frac{1}{ \det(\mathbf{h})}\left.\frac{ \partial E\left[\mathbf{r}, \mathbf{h}\left(\mathbf{1} + \bm{\epsilon}\right)\right]}{\partial \bm{\epsilon}} \right|_{\bm{\epsilon}=0}
\end{equation}

\begin{figure}
    \centering
    \includegraphics{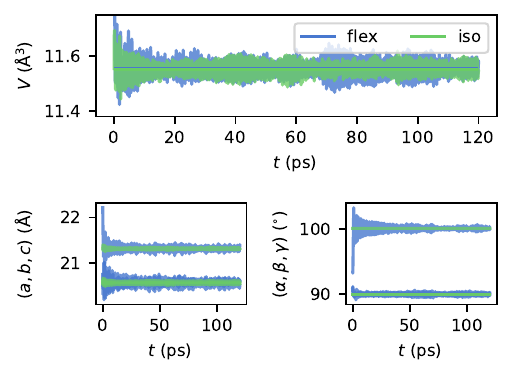}
    \caption{A comparison of the volume (top), lattice parameter (bottom left) and angle (bottom right) trajectory of an isotropic and a flexible-cell molecular dynamics run of \ce{HfO2} at T = \SI{298}{\kelvin}.}
    \label{fig:isovflex}
\end{figure}

The corresponding equations of motion describe the movement of the ions and their containing cell when interacting with a thermostat and barostat. To solve those equations, we resort to the Liouville factorization\cite{YU2010294} with a Nos\'e-Hoover chain coupling for both the thermostat and barostat. This code extension is also implemented on top of \textsc{jax}, rendering it fully automatically differentiable, and great care is taken to optimize the number of descriptor evaluations which constitute the bulk of run time, to optimize for efficiency.
To assess the integrity and correctness of the code, we benchmarked it against the preexisting isotropic implementation. As an example consider \cref{fig:isovflex}, where we show the volume, the lattice constants and the cell angles along the trajectory of a $768$-atom \ce{HfO2} system obtained using the flexible (blue) and the isotropic (green) barostat. For the flexible barostat, an arbitrary starting configuration is chosen, as can be seen by the rapidly adapting lattice constants in the panel in the bottom left. For the isotropic barostat, naturally, the proportion of the constants needs to be correct, as isotropic fluctuations cannot change it. Hence, the last \SI{50}{\pico\second} of the flexible-cell trajectory were averaged to obtain a reasonable starting point. The averages and standard deviations of the volume agree within less than \SI{1}{\percent} and \SI{20}{\percent}, respectively, between the isotropic and flexible simulation. The differences observed in the fluctuations are to be expected, as the constraint of isotropy leads to sampling a different distribution.

\section{Results and discussion}

\subsection{Volumetric analysis}
\begin{figure}
  \centering
  \includegraphics[width=\columnwidth]{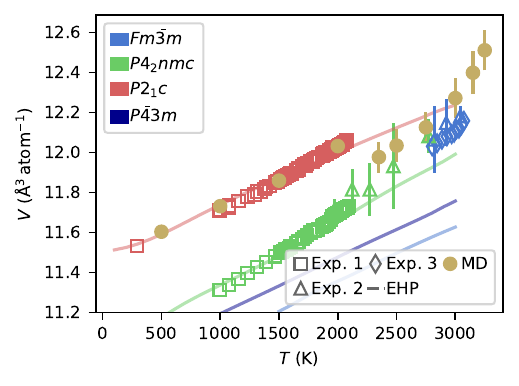}
  \caption{Volume as a function of temperature as obtained using MD (gold markers) compared with EHP results (solid lines) and several experimental investigations of different phases (Exp. 1: Ref.~\onlinecite{haggerty_jacs_14}, Exp. 2: Ref.~\onlinecite{tobase_pss_18}, Exp. 3: Ref.~\onlinecite{hong_nature_2018}) }
  \label{fig:volume}
\end{figure}
In \cref{fig:volume}, we compare the volumetric thermal expansion obtained using MD with the above-described methodology with experimental results \cite{haggerty_jacs_14,tobase_pss_18,hong_nature_2018} and our previous theoretical work carried out using EHPs \cite{bichelmaier_23b}. As mentioned earlier, the MD results are obtained under ambient pressure conditions (\SI{1}{\pbar}) and using a NNFF trained to parameterize the potential energy surface of the r$^2$SCAN functional. In the previous EHP work, the artifactual elongation of lattice constants introduced by the underlying PBE functional was compensated by the introduction of an artificial pressure of $P=\SI{4}{\giga\pascal}$ \cite{bichelmaier_23b}. We find that the r${}^{2}$SCAN NNFF reproduces the experimental volume very well without additional pressure and thus shows excellent predictive capabilities for \ce{HfO2}. In particular, the transition entailing a compression in volume is clearly visible around \SI{2000}{\kelvin} in accordance with the $P2_1/c$ to $P4_2/nmc$ transition \cite{wang_ACS_2006}. \ce{HfO2} is typically assumed to also exhibit a phase transition to a cubic structure in the temperature region above \SI{2500}{\kelvin}. However, the earlier EHP-predicted volumes obtained enforcing the cubic space groups $Fm\bar{3}m$ and $P\bar{4}3m$ substantially underestimate the cell volume, thereby calling this interpretation into question \cite{bichelmaier_23b}. It is noteworthy that the present volume predictions obtained from the MD trajectories are compatible with the available experimental data in the high-temperature region above \SI{2500}{\kelvin} even though no abrupt change in volume is predicted. 

\subsection{MD trajectory analysis}
\begin{figure}
  \centering
  \includegraphics[width=\columnwidth]{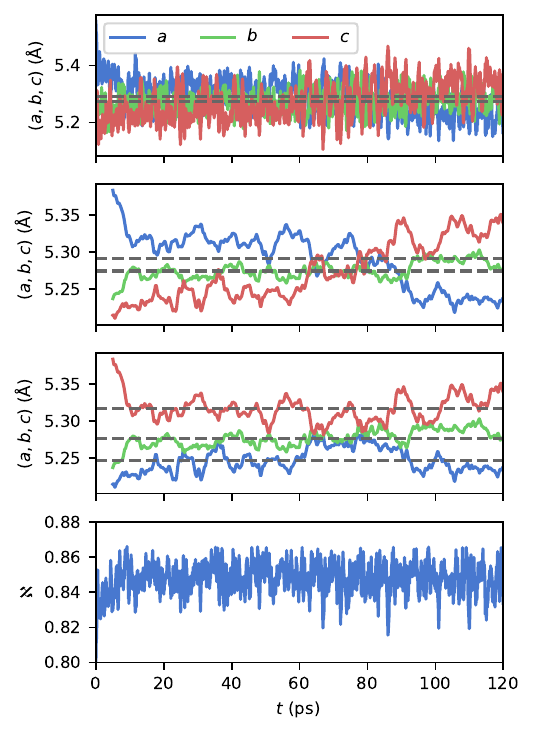}
  \caption{Lattice constants obtained from the $NPT$ MD trajectory at $P=\SI{1}{\bar}$ and \SI{3000}{\kelvin}. From top to bottom: (first) the raw trajectory data, (second) the result of applying a moving average across $\SI{500}{\femto\second}$, (third) the same, but after sorting the lattice parameters, and (fourth) the aspect ratio [$\aleph$; see \cref{eq:aleph}]. The horizontal dashed lines represent the respective averages.}
  \label{fig:t3000}
\end{figure}
To further investigate the high-temperature behavior of \ce{HfO2} we analyse the lattice constants throughout a $NPT$ MD trajectory. The lattice constants can undergo significant fluctuations during such runs, and the interpretation of their averages can be complicated or biased by the analysis procedure. Particularly in regions of metastability, the analysis of MD trajectories can be subject to ambiguity. As an example, consider the top panel of \cref{fig:t3000}, where a trajectory at \SI{3000}{\kelvin} is shown. Simply averaging over the trajectory of cell parameters, as is indicated by the horizontal dashed lines, might prompt the conclusion that the material is present in a cubic phase. However, the running average, depicted in the second panel, reveals that conclusion to be an artifact created by mixing together clearly non-cubic configurations where the elongated axis changes its orientation at around \SI{70}{\pico\second}. In the third panel of \cref{fig:t3000} the impact of that artifact is limited by sorting the running average of cell parameters at each timestep. 

\subsection{Aspect ratio}
\begin{figure}
  \centering
  \includegraphics{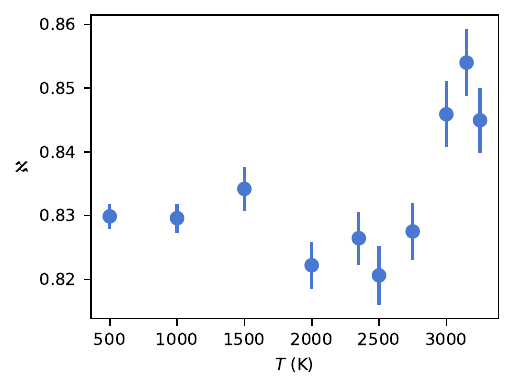}
  \caption[3D aspect ratio as a function of temperature]{Average aspect ratio ($\aleph$; see \cref{eq:aleph}) as a function of temperature. The error bars indicate the standard deviation along the trajectory.}  
  \label{fig:aleph}
\end{figure}
To better understand the behavior of the structure, a metric is required that does not depend on, e.g., the identification of the ``long'' axis. We thus introduce a three-dimensional aspect ratio, $\aleph$, defined as
\begin{equation}
  \aleph = \frac{\daleth}{\daleth_\mathrm{cub}}\text{, with }\daleth = \frac{V_i}{V_e},
  \label{eq:aleph}
\end{equation}
where $V_i$ ($V_e$) is the volume of the largest (smallest) sphere that can be inscribed (circumscribed) in the cell (i.e. $\daleth_\mathrm{cub}=\frac{1}{\sqrt{3}}$). We choose this normalization so that $\aleph$ attains its maximum value of $1$ if the structure is perfectly cubic. The aspect ratio obtained for the 3000~K trajectory is illustrated in the bottom panel of \cref{fig:t3000}. The value of $\aleph$ is seen to fluctuate around $0.845$ without being influenced by the change of orientation at 70~ps.

In \cref{fig:aleph} we show the value of $\aleph$ averaged over each trajectory as a function of temperature. First, a step-like change, larger than the standard deviation, is observed around \SI{2000}{\kelvin}, indicating the monoclinic to tetragonal phase transition taking place. At \SI{3000}{\kelvin} another step-like transition takes place indicating that the material is indeed increasing its cubicity. The temperature is in good agreement with the threshold often reported for a high-temperature phase transition in experimental studies. However, the $\aleph$ is still significantly smaller than one and, judging by the figure, it is highly unlikely that a cubic structure can be reached before melting. While a transition from monoclinic is clearly visible and agreement with literature is excellent, a transition to cubic is more ambiguous.

\subsection{Phase transitions}
\begin{figure}
    \centering
    \includegraphics{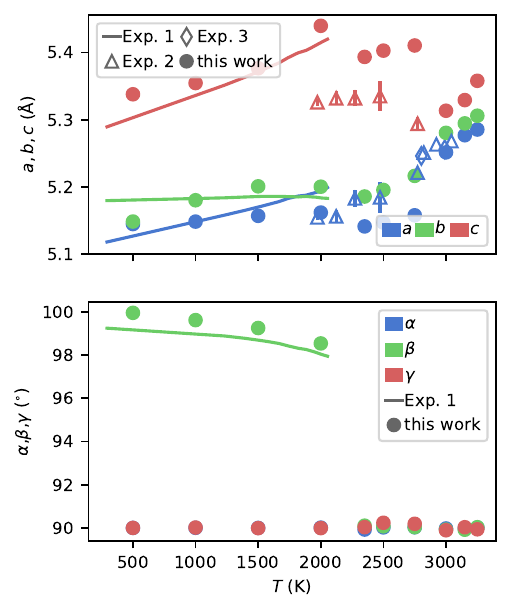} 
    \caption{Average lattice parameters as a function of temperature in comparison with experiment. (top) The lattice constants $(a,b,c)$ and (bottom) the angles $(\alpha,\beta,\gamma)$ as a function of temperature. The experimental values are taken from: Exp.~1: Ref.~\onlinecite{haggerty_jacs_14}, Exp.~2: Ref.~\onlinecite{tobase_pss_18}, Exp.~3: Ref.~\onlinecite{hong_nature_2018}. }
    \label{fig:lattice constants}
\end{figure}
To investigate the potential phase transitions in more detail, we show the lattice parameters as a function of temperature in \cref{fig:lattice constants}. Looking at the lattice constants in \cref{fig:lattice constants} we see an acceptable agreement with experimental results and a transition to an orthorhombic or tetragonal phase at \SI{2000}{\kelvin}. 
The angles $\alpha$ and $\gamma$ are oscillating around a mean of \SI{90}{\degree}, whereas the collapse of the angle $\beta$ from $\approx$ \SI{100}{\degree} to \SI{90}{\degree} at \SI{2000}{\kelvin} serves as clear indication for the transition away from monoclinic in very good agreement with the measurements by Haggerty et al. \cite{haggerty_jacs_14}. 

\begin{figure}
    \centering
    \includegraphics[width=\columnwidth]{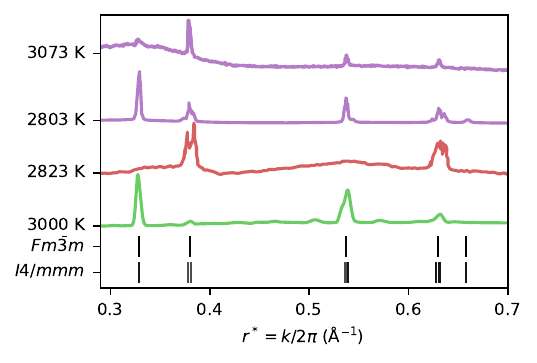}
    \caption{XRD patterns obtained from molecular dynamics (green) through \cref{eq:xrd}. Experimental data are taken from reports by Tobase et al. \cite{tobase_PRL_2018} (red) and Hong et al. \cite{hong_nature_2018} (violet) using WebPlotDigitizer \cite{Rohatgi2022}. Those experimental data are converted from $2\theta$ to $r^*$ using the reported energy (\SI{64.1}{\kilo\electronvolt}) \cite{tobase_pss_18} and wave length  ($\lambda$=\SI{0.124}{\angstrom}) \cite{hong_nature_2018} of the incident beams. The peak positions are obtained assuming lattice constants $a=\SI{5.263}{\angstrom}$ ($Fm\bar{3}m$)) and $a=\SI{3.7123}{\angstrom}$ and $c=\SI{5.290}{\angstrom}$ ($I4/mmm$)}
    \label{fig:xrd}
\end{figure}
The transition at around 3000~K that could be observed in the aspect ratio, \cref{fig:aleph}, can also be seen in the lattice constants [\cref{fig:lattice constants}, top]. However, they do not indicate a fully cubic phase and the transition is better described as one to a structure with a small tetragonal distortion.
The main experimental evidence for the high-temperature $Fm\bar{3}m$ \ce{HfO2} structure comes from X-ray diffraction patterns. \cref{fig:xrd} illustrates the patterns obtained in two recent experimental studies in red\cite{tobase_PRL_2018} and violet\cite{hong_nature_2018}. The peak positions are in good agreement with what would be expected for the $Fm\bar{3}m$ space group as shown by the black lines below. However, the extreme temperature means that only low angle peaks were observed. \cref{fig:aleph} and \cref{fig:lattice constants} would indicate that a small tetragonal distortion is still present at high temperature. The bottom of \cref{fig:xrd} shows that the peak positions of the $I4/mmm$ space group, which would result from a small tetragonal distortion of the $Fm\bar{3}m$ space group, give an equally good agreement with the available experimental data. To confirm that the MD trajectories are not in disagreement with the available high-temperature experimental evidence, we calculated the static structure factor from the \SI{3000}{\kelvin} trajectory according to \cref{eq:xrd}. This is shown with the green curve and a good agreement with the experimental peak positions can be observed. 

\section{Conclusion}
We have implemented a flexible-cell $NPT$ molecular dynamics workflow using automatic differentiation and trained a suitable neural-network force field on \ce{HfO2} data obtained using the r${}^{\mathrm{2}}$SCAN functional. These developments have been employed to perform molecular-dynamics simulations of \ce{HfO2} in the temperature range from \SI{500}{\kelvin} to \SI{3250}{\kelvin}. The lattice constants are in very good agreement with experiment and previous simulations, and clearly indicating a departure from monoclinic symmetry around \SI{2000}{\kelvin}. We have found no clear signs of a  transition towards any cubic phase at high temperatures. We find a small tetragonal distortion and show that the simulated X-ray diffraction patterns and lattice constants are still in good agreement with experimental data. We suggest $I4/mmm$ as a high-temperature space group in agreement with both simulated and experimental data.

\section*{Code availability}
The code implementing flexible-cell $NPT$ molecular dynamics on top of \textsc{jax-md} has been archived on Zenodo with D.O.I. \href{https://doi.org/10.5281/zenodo.10787829}{10.5281/zenodo.10787829}. A compatible version of \textsc{NeuralIL}, including example scripts for training and evaluation, has been archived on Zenodo with D.O.I. \href{https://doi.org/10.5281/zenodo.10786377}{10.5281/zenodo.10786377}. Both are distributed under an Apache 2 open source license.

\section*{Data availability}
A dataset containing the structures used for training and validation and their calculated energies and forces, the parameters of the neural-network force field, and a 3000~K MD trajectory are available on Zenodo with D.O.I. \href{https://doi.org/10.5281/zenodo.10793828}{10.5281/zenodo.10793828}.

\section*{Acknowledgements} \label{sec:acknowledgements}
AI4DI receives funding within the Electronic Components and Systems for European Leadership Joint Undertaking (ESCEL JU) in collaboration with the European Union’s Horizon2020 Framework Programme and National Authorities, under grant agreement n° 826060. This research was funded in part by the Austrian Science Fund (FWF) 10.55776/F81 (SFB F81 TACO).

%\bibliography{biblio.bib}
%merlin.mbs apsrev4-1.bst 2010-07-25 4.21a (PWD, AO, DPC) hacked
%Control: key (0)
%Control: author (72) initials jnrlst
%Control: editor formatted (1) identically to author
%Control: production of article title (-1) disabled
%Control: page (0) single
%Control: year (1) truncated
%Control: production of eprint (0) enabled
%

\end{document}